\documentstyle{article}

\begin{document}
\title{Energy levels of neutral atoms via a new perturbation method}
\author{Omar Mustafa$^{\dagger}$ and Maen Odeh$^{\ddagger}$ \\
Department of Physics, Eastern Mediterranean University\\
G. Magusa, North Cyprus, Mersin 10 - Turkey\\
$^{\dagger}$ omar.mustafa@mozart.emu.edu.tr\\
$^{\ddagger}$ odeh.as@mozart.emu.edu.tr\\
\date{}\\}
\maketitle
\begin{abstract}
{\small The energy levels of neutral atoms supported by Yukawa potential,
$V(r)=-Z exp(-\alpha r)/r$, are studied, using both dimensional and
dimensionless quantities, via a new analytical methodical proposal
( devised to solve for nonexactly solvable Schr\"odinger equation).
Using dimensionless quantities, by scaling the radial Hamiltonian
through $y=Zr$ and $\alpha^{'}=\alpha/Z$, we report that the scaled
screening parameter $\alpha^{'}$ is restricted to have values ranging
from zero to less than 0.4. On the other hand, working with the scaled
Hamiltonian enhances the accuracy and extremely speeds up the convergence
of the energy eigenvalues. The energy
levels of several new eligible scaled screening parameter $\alpha^{'}$
 values are also reported.}
\end{abstract}
\newpage

\section{Introduction}

The Yukawa potential plays special roles in different branches of physics.
It is known as the Debye-H\"uckel potential in plasma physics, as the
Thomas-Fermi or screened Coulomb potential in solid - state and atomic
physics, and the Yukawa potential in nuclear physics. It is also used in the
models describing metal - insulator transition [1,2]. As such, several
approaches have been invested to find the energy levels of this potential.
Amongst exist the simple variational procedure [3], the use of atomic
orbitals with a set of fitting parameters [4], Rayleigh - Schr\"odinger
perturbation theory [5,6], method of potential envelopes [7], iterative
procedure [8], different numerical calculations [9,10], and the shifted 1/N
expansion method [11] based on Papp's [12] or Imbo et al.'s [13,14]
proposals.

On the other hand, the Yukawa form\newline
\begin{equation}
V(r)=-Z exp(-\alpha r)/r
\end{equation}
\newline
is known to simulate the screening potential of neutral atoms, where the
screening parameter $\alpha$ is chosen to be $\alpha=\alpha_o Z^{1/3}$
corresponding to the Z - dependence of the reciprocal of the Thomas-Fermi
radius of the atom. However, invoking Fermi-Amaldi correction [15] in the
context of Ecker-Weizel approximation (EWA) method [16], Dutt and Varshni
[17] have amended the screening parameter and suggested that $%
\alpha=\alpha_o Z^{1/3}(1-1/Z)^{2/3}$ with $\alpha_o=0.98$. The correctness
of this amendment has been justified by Lai and Madan [18], who have used
the hypervirial - Pad\'{e} method (HPM) to obtain very accurate energy
eigenvalues. Therefore, the energy levels of the neutral atoms, supported by
potential (1), have been reexamined by applying the shifted large - N
expansion (SLNT) ( where N is the number of spatial dimensions) [14].
However, some drawbacks of the HPM and SLNT are in order. The HPM involves
elaborate computational time and effort for each numerical prediction, and
its application becomes quite restricted due to the absence of compact
analytical expressions for the eigenenergies and eigenfunctions. And,
because of the complexity in handling large - order corrections of the
standard Rayleigh - Schr\"odinger perturbation theory, only low - order
eigenenergy calculations have been reported for SLNT and large order -
calculations have been neglected [11-14]. Yet, neither SLNT nor HPM is
utilitarian in terms of calculating the eigenvalues and eigenfunctions in
one batch. Moreover, Scherrer et al. [19] have concluded, via matrix
continued fractions method, that SLNT leads to dubious accuracies. A
conclusion that we have very recently confirmed [20].

In this work we introduce a new methodical proposal for solving
Schr\"odinger equation with an arbitrary spherically symmetric potential. We
shall use $1/\bar{l}$ as an expansion parameter, where $\bar{l}=l-\beta$, $l$
is a quantum number and $\beta$ a suitable shift introduced, mainly, to a
avoid the trivial case $l=0$. In fact, the scope of its applicability is not
limited to the spherically symmetric problems, it extends also to systems at
lower dimensions. Here is the conversion strategy. One would rewrite the
centrifugal term $l(l+1)/2r^2$, in Eq.(2) below, as $\Lambda(%
\Lambda+1)/(2q^2)$ and use $1/\bar{\Lambda}$ as an expansion parameter. In
this case, $\Lambda=l$ in three dimensions, where $l$ is the angular
momentum quantum number with $q=r>0$, $\Lambda=|m|-1/2$ in two dimensions,
where m is the magnetic quantum number with $q=(x^2+y^2)^{1/2}>0$, and $%
\Lambda=-1$ and/or 0 in one dimension with $-\infty<q<\infty$. Thus the
method could be called a pseudoperturbative shifted - $l$ expansion
technique (PSLET), to be described in the following section.

\section{ The Method}

To construct the method we start with the time - independent one -
dimensional form of Schr\"odinger equation, in $\hbar = m = 1$ units,\newline
\begin{equation}
\left[-\frac{1}{2}\frac{d^{2}}{dr^{2}}+\frac{l(l+1)}{2r^{2}}+V(r)\right]
\Psi_{n_r,l}(r)=E_{n_r,l}\Psi_{n_r,l}(r).
\end{equation}
\newline
and to avoid the trivial case $l$=0, the quantum number $l$ is shifted
through the relation $\bar{l} = l - \beta$. Eq.(2) thus becomes\newline
\begin{equation}
\left\{-\frac{1}{2}\frac{d^{2}}{dr^{2}}+\tilde{V}(r)\right\} \Psi_{n_r,l}
(r)=E_{n_r,l}\Psi_{n_r,l}(r),
\end{equation}
\newline
\begin{equation}
\tilde{V}(r)=\frac{\bar{l}^{2}+(2\beta+1)\bar{l} +\beta(\beta+1)}{2r^{2}}+%
\frac{\bar{l}^2}{Q}V(r).
\end{equation}
\newline
Herein, it should be noted that Q is a constant that scales the potential $%
V(r)$ at large - $l$ limit and is set, for any specific choice of $l$ and $%
n_r$, equal to $\bar{l}^2$ at the end of the calculations [11-14,20-22].
And, $\beta$ is to be determined in the sequel.

Our systematic procedure begins with shifting the origin of the coordinate
through\newline
\begin{equation}
x=\bar{l}^{1/2}(r-r_{0})/r_{0},
\end{equation}
\newline
where $r_{0}$ is currently an arbitrary point to perform Taylor expansions
about, with its particular value to be determined. Expansions about this
point, $x=0$ (i.e. $r=r_0$), yield\newline
\begin{equation}
\frac{1}{r^{2}}=\sum^{\infty}_{n=0} (-1)^{n} \frac{(n+1)}{r_{0}^{2}} x^{n}%
\bar{l}^{-n/2},
\end{equation}
\newline
\begin{equation}
V(x(r))=\sum^{\infty}_{n=0}\left(\frac{d^{n}V(r_{0})}{dr_{0}^{n}}\right) 
\frac{(r_{0}x)^{n}}{n!}\bar{l}^{-n/2}.
\end{equation}
\newline
Obviously, the expansions in (6) and (7) center the problem at an arbitrary
point $r_0$ and the derivatives, in effect, contain information not only at $%
r_0$ but also at any point on the axis, in accordance with Taylor's theorem.
Also, it should be mentioned here that the scaled coordinate, equation (5),
has no effect on the energy eigenvalues, which are coordinate - independent.
It just facilitates the calculations of both the energy eigenvalues and
eigenfunctions. It is also convenient to expand $E$ as\newline
\begin{equation}
E_{n_r,l}=\sum^{\infty}_{n=-2}E_{n_r,l}^{(n)}\bar{l}^{-n}.
\end{equation}
\newline
Equation (3) thus becomes\newline
\begin{equation}
\left[-\frac{1}{2}\frac{d^{2}}{dx^{2}}+\frac{r_{0}^{2}}{\bar{l}} \tilde{V}%
(x(r))\right] \Psi_{n_r,l}(x)=\frac{r_{0}^2}{\bar{l}}E_{n_r,l}%
\Psi_{n_r,l}(x),
\end{equation}
\newline
with\newline
\begin{eqnarray}
\frac{r_0^2}{\bar{l}}\tilde{V}(x(r))&=&r_0^2\bar{l} \left[\frac{1}{2r_0^2}+%
\frac{V(r_0)}{Q}\right] +\bar{l}^{1/2}\left[-x+\frac{V^{^{\prime}}(r_0)r_0^3
x}{Q}\right]  \nonumber \\
&+&\left[\frac{3}{2}x^2+\frac{V^{^{\prime\prime}}(r_0) r_0^4 x^2}{2Q}\right]
+(2\beta+1)\sum^{\infty}_{n=1}(-1)^n \frac{(n+1)}{2}x^n \bar{l}^{-n/2} 
\nonumber \\
&+&r_0^2\sum^{\infty}_{n=3}\left[(-1)^n \frac{(n+1)}{2r_0^2}x^n +\left(\frac{%
d^n V(r_0)}{dr_0^n}\right)\frac{(r_0 x)^n}{n! Q}\right] \bar{l}^{-(n-2)/2} 
\nonumber \\
&+&\beta(\beta+1)\sum^{\infty}_{n=0}(-1)^n\frac{(n+1)}{2}x^n \bar{l}%
^{-(n+2)/2}+\frac{(2\beta+1)}{2},
\end{eqnarray}
\newline
where the prime of $V(r_0)$ denotes derivative with respect to $r_0$.
Equation (9) is exactly of the type of Schr\"odinger equation for one -
dimensional anharmonic oscillator\newline
\begin{equation}
\left[-\frac{1}{2}\frac{d^2}{dx^2}+\frac{1}{2}w^2 x^2 +\varepsilon_0 +P(x)%
\right]X_{n_r}(x)=\lambda_{n_r}X_{n_r}(x),
\end{equation}
\newline
where $P(x)$ is a perturbation - like term and $\varepsilon_0$ is a
constant. A simple comparison between Eqs.(9), (10) and (11) implies\newline
\begin{equation}
\varepsilon_0 =\bar{l}\left[\frac{1}{2}+\frac{r_0^2 V(r_0)}{Q}\right] +\frac{%
2\beta+1}{2}+\frac{\beta(\beta+1)}{2\bar{l}},
\end{equation}
\newline
\begin{eqnarray}
\lambda_{n_{r}}&=&\bar{l}\left[\frac{1}{2}+\frac{r_0^2 V(r_0)}{Q}\right] +%
\left[\frac{2\beta+1}{2}+(n_r+\frac{1}{2})w\right]  \nonumber \\
&+&\frac{1}{\bar{l}}\left[\frac{\beta(\beta+1)}{2}+\lambda_{n_{r}}^{(0)}%
\right] +\sum^{\infty}_{n=2}\lambda_{n_{r}}^{(n-1)}\bar{l}^{-n},
\end{eqnarray}
\newline
and\newline
\begin{equation}
\lambda_{n_{r}} = r_0^2 \sum^{\infty}_{n=-2} E_{n_r,l}^{(n)} \bar{l}%
^{-(n+1)},
\end{equation}
\newline
Equations (13) and (14) yield\newline
\begin{equation}
E_{n_r,l}^{(-2)}=\frac{1}{2r_0^2}+\frac{V(r_0)}{Q}
\end{equation}
\newline
\begin{equation}
E_{n_r,l}^{(-1)}=\frac{1}{r_0^2}\left[\frac{2\beta+1}{2} +(n_r +\frac{1}{2})w%
\right]
\end{equation}
\newline
\begin{equation}
E_{n_r,l}^{(0)}=\frac{1}{r_0^2}\left[ \frac{\beta(\beta+1)}{2}
+\lambda_{n_r}^{(0)}\right]
\end{equation}
\newline
\begin{equation}
E_{n_r,l}^{(n)}=\lambda_{n_r}^{(n)}/r_0^2 ~~;~~~~n \geq 1.
\end{equation}
\newline
Here $r_0$ is chosen to minimize $E_{n_r,l}^{(-2)}$, i. e.\newline
\begin{equation}
\frac{dE_{n_r,l}^{(-2)}}{dr_0}=0~~~~ and~~~~\frac{d^2 E_{n_r,l}^{(-2)}}{%
dr_0^2}>0.
\end{equation}
\newline
Hereby, $V(r)$ is assumed to be well behaved so that $E^{(-2)}$ has a
minimum $r_0$ and there are well - defined bound - states. Equation (19) in
turn gives, with $\bar{l}=\sqrt{Q}$,\newline
\begin{equation}
l-\beta=\sqrt{r_{0}^{3}V^{^{\prime}}(r_{0})}.
\end{equation}
\newline
Consequently, the second term in Eq.(10) vanishes and the first term adds a
constant to the energy eigenvalues. It should be noted that energy term $%
\bar{l}^2E_{n_r,l}^{(-2)}$ has its counterpart in classical mechanics. It
corresponds roughly to the energy of a classical particle with angular
momentum $L_z$=$\bar{l}$ executing circular motion of radius $r_0$ in the
potential $V(r_0)$. This term thus identifies the leading - order
approximation, to all eigenvalues, as a classical approximation and the
higher - order corrections as quantum fluctuations around the minimum $r_0$,
organized in inverse powers of $\bar{l}$.

The next leading correction to the energy series, $\bar{l}E_{n_r,l}^{(-1)}$,
consists of a constant term and the exact eigenvalues of the unperturbed
harmonic oscillator potential $w^2x^2/2$. The shifting parameter $\beta$ is
determined by choosing $\bar{l}E_{n_r,l}^{(-1)}$=0. This choice is
physically motivated. It requires not only the agreements between PSLET
eigenvalues and the exact known ones for the harmonic oscillator and Coulomb
potentials but also between the eigenfunctions as well. Hence\newline
\begin{equation}
\beta=-\left[\frac{1}{2}+(n_{r}+\frac{1}{2})w\right],
\end{equation}
\newline
where\newline
\begin{equation}
w=\sqrt{3+\frac{r_0 V^{^{\prime\prime}}(r_0)}{V^{^{\prime}}(r_0)}}.
\end{equation}
\newline

Then equation (10) reduces to\newline
\begin{equation}
\frac{r_0^2}{\bar{l}}\tilde{V}(x(r))= r_0^2\bar{l}\left[\frac{1}{2r_0^2}+%
\frac{V(r_0)}{Q}\right]+ \sum^{\infty}_{n=0} v^{(n)}(x) \bar{l}^{-n/2},
\end{equation}
\newline
where\newline
\begin{equation}
v^{(0)}(x)=\frac{1}{2}w^2 x^2 + \frac{2\beta+1}{2},
\end{equation}
\newline
\begin{equation}
v^{(1)}(x)=-(2\beta+1) x - 2x^3 + \frac{r_0^5 V^{^{\prime\prime\prime}}(r_0)%
}{6 Q} x^3,
\end{equation}
\newline
and for $n \geq 2$\newline
\begin{eqnarray}
v^{(n)}(x)&=&(-1)^n (2\beta+1) \frac{(n+1)}{2} x^n + (-1)^{n} \frac{%
\beta(\beta+1)}{2} (n-1) x^{(n-2)}  \nonumber \\
&+& \left[(-1)^{n} \frac{(n+3)}{2} + \frac{r_0^{(n+4)}}{Q(n+2)!} \frac{%
d^{n+2} V(r_0)}{dr_0^{n+2}}\right] x^{n+2}.
\end{eqnarray}
\newline
Equation (9) thus becomes\newline
\begin{eqnarray}
&&\left[-\frac{1}{2}\frac{d^2}{dx^2} + \sum^{\infty}_{n=0} v^{(n)} \bar{l}%
^{-n/2}\right]\Psi_{n_r,l} (x)=  \nonumber \\
&& \left[\frac{1}{\bar{l}}\left(\frac{\beta(\beta+1)}{2} +%
\lambda_{n_r}^{(0)}\right) + \sum^{\infty}_{n=2} \lambda_{n_r}^{(n-1)} \bar{l%
}^{-n} \right] \Psi_{n_r,l}(x).
\end{eqnarray}
\newline

When setting the nodeless, $n_r = 0$, wave functions as \newline
\begin{equation}
\Psi_{0,l}(x(r)) = exp(U_{0,l}(x)),
\end{equation}
\newline
equation (27) is readily transformed into the following Riccati equation:%
\newline
\begin{eqnarray}
-\frac{1}{2}[ U^{^{\prime\prime}}(x)+U^{^{\prime}}(x)U^{^{\prime}}(x)]
+\sum^{\infty}_{n=0} v^{(n)}(x) \bar{l}^{-n/2} &=&\frac{1}{\bar{l}} \left( 
\frac{\beta(\beta+1)}{2} + \lambda_{0}^{(0)}\right)  \nonumber \\
&&+\sum^{\infty}_{n=2} \lambda_{0}^{(n-1)} \bar{l}^{-n}.
\end{eqnarray}
\newline
Hereinafter, we shall use $U(x)$ instead of $U_{0,l}(x)$ for simplicity, and
the prime of $U(x)$ denotes derivative with respect to $x$. It is evident
that this equation admits solution of the form \newline
\begin{equation}
U^{^{\prime}}(x)=\sum^{\infty}_{n=0} U^{(n)}(x) \bar{l}^{-n/2}
+\sum^{\infty}_{n=0} G^{(n)}(x) \bar{l}^{-(n+1)/2},
\end{equation}
\newline
where\newline
\begin{equation}
U^{(n)}(x)=\sum^{n+1}_{m=0} D_{m,n} x^{2m-1} ~~~~;~~~D_{0,n}=0,
\end{equation}
\newline
\begin{equation}
G^{(n)}(x)=\sum^{n+1}_{m=0} C_{m,n} x^{2m}.
\end{equation}
\newline
Substituting equations (30) - (32) into equation (29) implies\newline
\begin{eqnarray}
&-&\frac{1}{2} \sum^{\infty}_{n=0}\left[U^{(n)^{^{\prime}}} \bar{l}^{-n/2} +
G^{(n)^{^{\prime}}} \bar{l}^{-(n+1)/2}\right]  \nonumber \\
&-&\frac{1}{2} \sum^{\infty}_{n=0} \sum^{\infty}_{p=0} \left[ U^{(n)}U^{(p)} 
\bar{l}^{-(n+p)/2} +G^{(n)}G^{(p)} \bar{l}^{-(n+p+2)/2} +2 U^{(n)}G^{(p)} 
\bar{l}^{-(n+p+1)/2}\right]  \nonumber \\
&+&\sum^{\infty}_{n=0}v^{(n)} \bar{l}^{-n/2} =\frac{1}{\bar{l}}\left(\frac{%
\beta(\beta+1)}{2}+\lambda_{0}^{(0)}\right) +\sum^{\infty}_{n=2}
\lambda_{0}^{(n-1)} \bar{l}^{-n},
\end{eqnarray}
\newline
where primes of $U^{(n)}(x)$ and $G^{(n)}(x)$ denote derivatives with
respect to $x$. Equating the coefficients of the same powers of $\bar{l}$
and $x$, respectively, ( of course the other way around would work equally
well) one obtains\newline
\begin{equation}
-\frac{1}{2}U^{(0)^{^{\prime}}} - \frac{1}{2} U^{(0)} U^{(0)} + v^{(0)} = 0,
\end{equation}
\newline
\begin{equation}
U^{(0)^{^{\prime}}}(x) = D_{1,0} ~~~;~~~~D_{1,0}=-w,
\end{equation}
\newline
and integration over $dx$ yields\newline
\begin{equation}
U^{(0)}(x)=-wx.
\end{equation}
\newline
Similarly,\newline
\begin{equation}
-\frac{1}{2}[U^{(1)^{^{\prime}}} + G^{(0)^{^{\prime}}}] - U^{(0)}U^{(1)} -
U^{(0)}G^{(0)} +v^{(1)}=0,
\end{equation}
\newline
\begin{equation}
U^{(1)}(x)=0,
\end{equation}
\newline
\begin{equation}
G^{(0)}(x)=C_{0,0}+C_{1,0}x^2,
\end{equation}
\newline
\begin{equation}
C_{1,0}=-\frac{B_{1}}{w},
\end{equation}
\newline
\begin{equation}
C_{0,0}=\frac{1}{w}(C_{1,0}+2\beta+1),
\end{equation}
\newline
\begin{equation}
B_{1}=-2+\frac{r_0^5}{6Q}\frac{d^3 V(r_0)}{dr_0^3},
\end{equation}
\newline
\begin{eqnarray}
&&-\frac{1}{2}[U^{(2)^{^{\prime}}} + G^{(1)^{^{\prime}}}] - \frac{1}{2}%
\sum^{2}_{n=0}U^{(n)}U^{(2-n)}-\frac{1}{2}G^{(0)}G^{(0)}  \nonumber \\
&&-\sum^{1}_{n=0}U^{(n)}G^{(1-n)} + v^{(2)} = \frac{\beta(\beta+1)}{2} +
\lambda_{0}^{(0)},
\end{eqnarray}
\newline
\begin{equation}
U^{(2)}(x)=D_{1,2}x + D_{2,2}x^3,
\end{equation}
\newline
\begin{equation}
G^{(1)}(x)=0,
\end{equation}
\newline
\begin{equation}
D_{2,2}=\frac{1}{w}(\frac{C_{1,0}^2}{2}-B_{2})
\end{equation}
\newline
\begin{equation}
D_{1,2}=\frac{1}{w}(\frac{3}{2}D_{2,2}+C_{0,0}C_{1,0} -\frac{3}{2}%
(2\beta+1)),
\end{equation}
\newline
\begin{equation}
B_{2}=\frac{5}{2}+\frac{r_0^6}{24Q}\frac{d^4V(r_0)}{dr_0^4},
\end{equation}
\newline
\begin{equation}
\lambda_{0}^{(0)} = -\frac{1}{2}(D_{1,2}+C_{0,0}^2).
\end{equation}
\newline
$\cdots$ and so on. Thus, one can calculate the energy eigenvalue and the
eigenfunctions from the knowledge of $C_{m,n}$ and $D_{m,n}$ in a
hierarchical manner. Nevertheless, the procedure just described is suitable
for systematic calculations using software packages (such as MATHEMATICA,
MAPLE, or REDUCE) to determine the energy eigenvalue and eigenfunction
corrections up to any order of the pseudoperturbation series.

Although the energy series, Eq.(8), could appear divergent, or, at best,
asymptotic for small $\bar{l}$, one can still calculate the eigenenergies to
a very good accuracy by forming the sophisticated Pade' approximation to the
energy series. Our strategy is therefore clear.

\section{Yukawa Potential}

Let us begin with the Yukawa potential (1), where the screening parameter $%
\alpha$ is given by the amended Z-dependent relation\newline
\begin{equation}
\alpha=0.98 Z^{1/3} (1-1/Z)^{2/3}.
\end{equation}
\newline
Thus, Eq.(22) with $n_r=0$ reads\newline
\begin{equation}
w=\sqrt{\frac{-\alpha^2r_0^2+\alpha r_0+1}{\alpha r_0+1}},
\end{equation}
\newline
and Eq.(20) in turn implies\newline
\begin{equation}
l+\frac{1}{2}(1+w)=\sqrt{Z r_0 e^{-\alpha r_0} (\alpha r_0+1)}.
\end{equation}
\newline
Evidently, one should appeal to numerical techniques to solve for $r_0$
since a closed form solution for Eq.(52) is hard to find, if it is feasible
at all. Once $r_0$ is determined the coefficients $C_{m,n}$ and $D_{m,n}$
are obtained in a sequel manner. Consequently, the eigenvalues and
eigengfunctions are calculated in the same batch for each value of $Z$ and $%
\alpha$.

In tables 1 and 2, we list our K- and L-shell binding energies, measured in
units of 2Ry=27.196 eV, for some values of $Z$ and compare with those of the
hypervirial - Pad\'{e} method (HPM) [18], and of the shifted large - N
expansion technique (SLNT) [14]. In table 1, the results of Ecker - Weizel
approximation (EWA) [16] are also displayed. The first ten terms of our
energy series are calculated and the stability of the sequence of Pad\'{e}
approximants is considered.

Evidently, our results compare better with those of HPM than the results of
SLNT. However, considering only the first four terms of our series, our
results are found in exact agreement with those of SLNT. The accuracy of our
results is enhanced by the Pad\'{e} approximants which show stability all
the way from $E[2,2]$ to $E[4,5]$. Such stability indicates that our results
are exact, provided that some uncertainty lies in the last $j$ digits in
parentheses.

On the other hand, one would pass to dimensionless quantities by scaling the
Hamiltonian and reducing it to the form\newline
\begin{equation}
H_y=-\frac{1}{2}\frac{d^2}{dy^2}+\frac{l(l+1)}{2y^2} -\frac{1}{y}%
e^{-\alpha^{\prime}y},
\end{equation}
\newline
where $y\in(0,\infty)$ and $\alpha^{\prime}$ is a positive screening
parameter. Then, to return back to dimensional quantities one has to make
the following substitutions\newline
\begin{equation}
y=Zr~~~,~~~\alpha^{\prime}=\frac{\alpha}{Z}=0.98 Z^{-2/3} (1-\frac{1}{Z}%
)^{2/3},
\end{equation}
\newline
where $Z=1,2,3,\cdots$. Therefore, $\alpha^{\prime}$ can not be a random
screening parameter. It is restricted by Eq.(54), which in effect implies
that the lowest value of $\alpha^{\prime}$ is zero when $Z=1$ and
immediately a maximum value 0.38891326 at $Z$=2. For $Z>2$, $\alpha^{\prime}$
decreases smoothly and converges to zero as $Z\rightarrow\infty$. To figure
this out, one should simply sketch the relation of $\alpha^{\prime}$ vs $Z$,
given in Eq.(54). Hence, we proceed with our comparison, using only several
eligible values for $\alpha^{\prime}$, and list our results (tables 3 and 4)
along with those of Varshni [11],  via SLNT, Fernandez et al. [23],  via a
rational function approximation (RFA), and the exact ones [24,25]. Our
results, again, are in better agreement with the RFA and the exact ones than
those of Varshni [11], via Papp's [12] or Imbo's [13] SLNT proposals.

Although we have used the atomic units through out, our energies in tables
1 and 2 are multiplied by 2Ry for the sake of comparison with the results
from HPM [18], EWA [16], and SLNT [14].

\section{Concluding remarks}

In view of the results reported in tables 1-4, some remarks merit to be
mentioned.

It is evident that working with the scaled Hamiltonian , Eq.(53),
facilitates the calculations and enhances the accuracy of PSLET. As can be
seen from our results, the convergence for the energy eigenvalues is
extremely rapid when the scaled Hamiltonian is used. It should be obvious
for the reader to see that all $Z$ - values in tables 1-2 fall in the range
of $\alpha^{\prime}$ in tables 3 and 4. Moreover, as $\alpha^{\prime}$ gets
smaller the Yukawa potential becomes more Coulombic in nature, and when $%
\alpha^{\prime}$=0 at $Z$=1 the Yukawa collapses into the exact Coulombic
potential, for which PSLET gives exact spectrum from the leading term $\bar{l%
}^2E^{(-2)}$ and higher - order corrections vanish identically.

The answer to the question as to why all methods reported by Varshni [11]
give poor accuracies for $\alpha^{\prime}\geq 0.7$, lies perhaps in the
restriction provided by Eq.(54) that $\alpha^{\prime}< 0.4$ (table 1 of
ref.[11] bears this out). Moreover, the non-negligible deviations of SLNT
results [19,20,26-30] from the exact values should be attributed, mainly, to
the limited capability of SLNT to calculate the energy corrections beyond
the fourth - order.

From the knowledge of $C_{m,n}$ and $D_{m,n}$ one can calculate, in the same
batch, the wave functions to study electronic transitions and multiphoton
emission occurring in atomic systems in the presence of intense laser
fields, for example. Such studies already lie beyond the scope of our
present methodical proposal.

To sum up, we have demonstrated that it is an easy task to implement PSLET
without having to worry about the ranges of couplings and forms of
perturbations in the potential involved. In contrast to the textbook
Rayleigh - Schr\"odinger perturbation theory, an easy feasibility of
computation of the eigenvalues and eigenfunctions, in one batch, has been
demonstrated, and satisfactory accuracies have been obtained. Moreover, a
nice numerical trend of convergence has been achieved. Nevertheless, another
suitable criterion for choosing the value of the shift $\beta$, reported in
ref. [30], is also feasible. This reference should be consulted for more
details.

\newpage 
\begin{table}[tbp]
\caption{ K-shell energies, with $n_r=0$ and $l=0$, in KeV for $V(r)=-Ze^{-%
\protect\alpha r}/r$. Where $E_{PSLET}$ represents PSLET results, $E_{HPM}$
denotes the Hypervirial method [18], $E_{EWA}$ from [16], and $E_{SLNT}$
from [14]. $E[4,5]$ is the [4,5] Pad\'{e} approximant obtained by replacing
the last $j$ digits of $E[4,4]$ with the $j$ digits in parentheses.}
\begin{center}
\vspace{1cm} 
\begin{tabular}{|cccccc|}
\hline\hline
$Z$ & $-E_{EWA}$ & $-E_{HPM}$ & $-E_{SLNT}$ & $-E_{PSLET}$ & $%
-E[4,4]~\&~(-E[4,5])$ \\ \hline
3 & 0.05334 & 0.05415 & 0.05414 & 0.054135 & 0.05414708 (19) \\ 
6 & 0.274 & 0.27623 & 0.27623 & 0.276226 & 0.27622917 (22) \\ 
9 & 0.701 & 0.70461 & 0.70461 & 0.704612 & 0.70461288 (9) \\ 
14 & 1.897 & 1.90320 & 1.90320 & 1.903198 & 1.903198386 (6) \\ 
19 & 3.716 & 3.72545 & 3.72545 & 3.725447 & 3.725447269 (7) \\ 
24 & 6.171 & 6.18277 & 6.18277 & 6.182765 & 6.182765210 (07) \\ 
29 & 9.268 & 9.28212 & 9.28213 & 9.282124 & 9.282124421 (18) \\ 
34 & 13.012 & 13.02830 & 13.02830 & 13.028304 & 13.028303677 (4) \\ 
39 & 17.407 & 17.42482 & 17.42482 & 17.424818 & 17.424817591 (88) \\ 
44 & 22.454 & 22.47438 & 22.47438 & 22.474378 & 22.474378305 (3) \\ 
49 & 28.157 & 28.17915 & 28.17915 & 28.179153 & 28.179153161 (0) \\ 
54 & 34.517 & 34.54092 & 34.54092 & 34.540921 & 34.540920826 (5) \\ 
59 & 41.535 & 41.56612 & 41.56117 & 41.56117189 & 41.561171872 (1) \\ 
64 & 49.213 & 49.24118 & 49.24118 & 49.2411768 & 49.241176745 (4) \\ 
69 & 57.553 & 57.58203 & 57.58203 & 57.5820335 & 57.58203349 (8) \\ 
74 & 66.554 & 66.58470 & 66.58470 & 66.5847023 & 66.584702299 (9) \\ 
79 & 76.217 & 76.25003 & 76.25003 & 76.2500313 & 76.250031268 (8) \\ 
84 & 86.544 & 86.57878 & 86.57878 & 86.5787759 & 86.578775902 (1) \\ 
\hline\hline
\end{tabular}
\end{center}
\end{table}
\newpage 
\begin{table}[tbp]
\caption{ L-shell energies, with $n_r=0$ and $l=1$, in KeV for $V(r)=-Ze^{-%
\protect\alpha r}/r$. Where $E_{PSLET}$ represents PSLET results, $E_{HPM}$
denotes the Hypervirial method [18], and $E_{SLNT}$ from [14]. $E[4,5]$ is
the [4,5] Pad\'{e} approximant obtained by replacing the last $j$ digits of $%
E[4,4]$ with the $j$ digits in parentheses.}
\begin{center}
\vspace{1cm} 
\begin{tabular}{|ccccc|}
\hline\hline
$Z$ & $-E_{HPM}$ & $-E_{SLNT}$ & $-E_{PSLET}$ & $-E[4,4]~\&~(-E[4,5])$ \\ 
\hline
9 & 0.00423 & 0.00437 & 0.0056 & 0.0041 (1) \\ 
14 & 0.08953 & 0.08945 & 0.08948 & 0.089523 (31) \\ 
19 & 0.28398 & 0.28394 & 0.283962 & 0.283977 (8) \\ 
24 & 0.59991 & 0.59989 & 0.599905 & 0.599912 (2) \\ 
29 & 1.04510 & 1.04508 & 0.045092 & 1.0450957 (8) \\ 
34 & 1.62485 & 1.62484 & 1.624851 & 1.6248535 (6) \\ 
39 & 2.34309 & 2.34308 & 2.343087 & 2.3430888 (9) \\ 
44 & 3.20280 & 3.20280 & 3.203800 & 3.2028012 (3) \\ 
49 & 4.20638 & 4.20637 & 4.2063780 & 4.20637855 (9) \\ 
54 & 5.35577 & 5.35577 & 5.3557723 & 5.35577266 (9) \\ 
59 & 6.65261 & 6.65261 & 6.6526133 & 6.65261354 (6) \\ 
64 & 8.09829 & 8.09828 & 8.0982857 & 8.09828593 (5) \\ 
69 & 9.69398 & 9.69398 & 9.6939829 & 9.69398303 (4) \\ 
74 & 11.44075 & 11.44074 & 11.4407452 & 11.44074536 (7) \\ 
79 & 13.33949 & 13.33949 & 13.3394894 & 13.33948956 (7) \\ 
84 & 15.39103 & 15.39103 & 15.3910302 & 15.39103029 (30) \\ \hline\hline
\end{tabular}
\end{center}
\end{table}
\newpage 
\begin{table}[tbp]
\caption{ K-shell energies, with $n_r=0$ and $l=0$, in $\hbar=m=e=1$ units
for $V(r)=-e^{-\protect\alpha^{\prime}r}/r$. Where $E_{PSLET}$ represents
PSLET results, $E_{Papp}$ denotes SLNT, via Papps proposal [11],and $E_{Imbo}
$ SLNT, via Imbo's proposal [11]. $E[4,5]$ is the [4,5] Pad\'{e} approximant
obtained by replacing the last $j$ digits of $E[4,4]$ with the $j$ digits in
parentheses.}
\begin{center}
\vspace{1cm} 
\begin{tabular}{|cccccc|}
\hline\hline
$\alpha^{\prime}$ & $-E_{Papp}$ & $-E_{Imbo}$ & $-E_{PSLET}$ & $%
-E[4,4]~\&~(-E[4,5])$ & Exact [23-25] \\ \hline
0.1 & 0.407058 & 0.407058 & 0.40705803 & 0.407058031 (1) & 0.407058031 \\ 
0.2 & 0.326808 & 0.326808 & 0.326808231 & 0.326808515 (9) & 0.326808515 \\ 
0.3 & 0.257634 & 0.257634 & 0.257628168 & 0.25763869 (8) & 0.25763869 \\ 
0.4 & 0.198346 & 0.198345 & 0.198260722 & 0.198377 (8) & 0.198377 \\ 
\hline\hline
\end{tabular}
\end{center}
\end{table}
\begin{table}[tbp]
\caption{ K-shell energies, with $n_r=0$ and $l=0$, in $\hbar=m=e=1$ units
for $V(r)=-e^{-\protect\alpha^{\prime}r}/r$. Where $E_{PSLET}$ represents
PSLET results, $E[4,5]$ is the [4,5] Pad\'{e} approximant obtained by
replacing the last $j$ digits of $E[4,4]$ with the $j$ digits in parentheses.
}
\begin{center}
\vspace{1cm} 
\begin{tabular}{|ccc|}
\hline\hline
$\alpha^{\prime}$ & $-E_{PSLET}$ & $-E[4,4]~\&~(-E[4,5])$ \\ \hline
0.01 & 0.4900745067 & 0.490074506746694 (4) \\ 
0.02 & 0.48029610598 & 0.48029610598378 (8) \\ 
0.03 & 0.4706620270266 & 0.4706620270246 (4) \\ \hline\hline
\end{tabular}
\end{center}
\end{table}

\end{document}